\documentclass[conference]{IEEEtran}
\IEEEoverridecommandlockouts
\usepackage{cite}
\usepackage{amsmath,amssymb,amsfonts}
\usepackage{algorithmic}
\usepackage{graphicx}
\usepackage{textcomp}
\usepackage{xcolor}
\def\BibTeX{{\rm B\kern-.05em{\sc i\kern-.025em b}\kern-.08em
    T\kern-.1667em\lower.7ex\hbox{E}\kern-.125emX}}
\begin{document}

\title{Can Large Language Models Simulate Symbolic Execution Output Like KLEE?\\
}

\author{\IEEEauthorblockN{Rong Feng, Vanisha Gupta, Vivek Patel, Viroopaksh Reddy Ernampati, Suman Saha  }
\IEEEauthorblockA{\textit{Dept. of Computer Science and Engineering} \\
\textit{The Pennsylvania State University)}\\
University Park, PA, USA \\
\{rjf5768, vqg5208, vmp5428, vze5037\}@psu.edu}


}

\maketitle

\begin{abstract}
Symbolic execution helps check programs by exploring different paths based on symbolic inputs. Tools like KLEE are commonly used because they can automatically detect bugs and create test cases. But one of KLEE’s biggest issues is how slow it can get when programs have lots of branching paths—it often becomes too resource-heavy to run on large or complex code. In this project, we wanted to see if a large language model like GPT-4o could simulate the kinds of outputs that KLEE generates. The idea was to explore whether LLMs could one day replace parts of symbolic execution to save time and resources.

One specific goal was to have GPT-4o identify the most constrained path in a program—this is the execution path with the most symbolic conditions. These paths are especially important because they often represent edge cases that are harder to test and more likely to contain deep bugs. However, figuring this out usually requires fully running KLEE, which can be expensive. So, we tested whether GPT-4o could predict the KLEE outputs and the most complex path using a dataset of 100 C programs. Our results showed about 20\% accuracy in generating KLEE-like outputs and identifying the most constrained path. While not highly accurate, this early work helps show what current LLMs can and can’t do when it comes to simulating symbolic execution.

.
\end{abstract}


\section{Introduction}

Symbolic execution is a method used to check how programs behave by exploring different execution paths using symbolic inputs instead of real values. This lets tools test many possible input combinations to uncover edge cases or hidden bugs that might not show up with regular testing \cite{cadar2013symbolic}. It’s beneficial for automatically creating test cases, detecting bugs, and verifying whether a program works as expected \cite{uehara2016exhaustive, brown2020sys, ibing2015symbolic}. One of the most popular tools that use symbolic execution is KLEE \cite{cadar2008klee}. While KLEE is very effective, it struggles when programs have a lot of conditional branches or get too large. This is because of something called the path explosion problem—where the number of possible paths grows super fast as programs get more complex \cite{bessler2021metrinome}. In those situations, running complete symbolic execution with KLEE can take too much time and memory, which makes it hard for developers to use it on large-scale projects.

Another challenge in symbolic execution, especially when using tools like KLEE, is identifying what’s known as the \textit{most constrained execution path}. This path in the program has the highest number of symbolic conditions, which usually makes it the most complex and challenging to satisfy \cite{engler2007under}. These paths are critical because they often point to rare edge cases where bugs are more likely to hide or the program behaves unexpectedly. By focusing on the most constrained path, testers can prioritize the part of the code that’s hardest to reason about and most likely to cause issues. The problem is that finding this path isn’t easy. It requires running symbolic execution across all possible paths, which takes time and memory—especially for programs with lots of branches or inputs \cite{kapus2020pending}. Although KLEE does produce output that includes this information, getting to the most constrained path still means analyzing hundreds of test cases or constraint files, which isn’t practical at scale.

Given how slow and resource-heavy complete symbolic execution can be, it’s natural to wonder if large language models (LLMs) could help with some of that work. In recent years, models like GPT-4 have shown strong results in software development tasks. They’ve been used to write code, generate test cases, detect bugs, and assist in analyzing programs \cite{ibrahimzada2023automated, martin2008automatic, kang2023large}. More recently, researchers have begun exploring how LLMs might be used to support symbolic execution—for example, by generating inputs that explore more paths, reducing false positives, or guiding analysis toward potentially buggy code \cite{nguyen2024generate, pan2024multi, sen2024sellm, zaharudin2024enhancing, xu2024symbolic}. Unlike traditional tools, LLMs can learn patterns from previous examples and apply that knowledge to new code. This ability to generalize could help make program analysis faster and more scalable, especially if they can take over some of the heavy lifting done by tools like KLEE.

In this project, we wanted to determine if a language model like GPT-4o could act as a lighter and faster alternative to KLEE by generating similar outputs without actually running symbolic execution. Since KLEE can be slow and resource-heavy, especially on programs with many branches, we were curious whether an LLM could learn the patterns behind symbolic outputs and predict them directly from the code. We fine-tuned GPT-4o using 80 C programs and their corresponding files generated by KLEE. Then, we tested the model on 20 new programs it hadn’t seen before. The goal was to determine whether GPT-4o could not only generate outputs similar to KLEE’s, but also correctly identify the most constrained execution path. While the model didn’t perform very well—it matched KLEE’s outputs and identified the correct path only about 20\% of the time—this still gave us useful insight into what LLMs can and can’t do. Our results suggest that current language models aren’t yet ready to replace symbolic execution tools, but they might be able to support or partially simulate them in the future.

\section{Data Collection and Processing}

In this section, we explain how we collected and prepared C programs compatible with KLEE to generate the training data needed to fine-tune GPT-4o. We also describe how we used KLEE’s output to identify the most constrained execution path, which we later included as part of the model’s training.

\subsection{Data Collection}

To build our dataset, we started by searching for open-source C programs that would work well with KLEE. After looking through a few options, we chose the TheAlgorithms/C GitHub repository \cite{TheAlgorithmsC} because it contains a large number of clean, standalone algorithm implementations. These programs are simple, self-contained, and easy to run, which makes them a good fit for symbolic execution.
We followed strict rules to ensure KLEE could analyze each program without issues. First, every program had to have its \texttt{main()} function since that’s where KLEE begins symbolic execution. We avoided programs that depended on external libraries or multiple files because those often required extra setup. 
Additionally, we excluded programs that used input functions like \texttt{scanf()} or \texttt{fgets()} because KLEE needs input to be symbolic rather than read from the user. We wrote a Python script to automatically apply these filters, and after running it, we ended up with about 100 programs that were fully compatible and ready for use in our next phase of experiments.


\subsection{Data Processing}
To symbolically execute each C program, we followed a few straightforward steps. First, we modified the source code to include special KLEE functions like \texttt{klee\_make\_symbolic(\&x, sizeof(x), "x");}. This tells KLEE which variables should be treated as symbolic, allowing it to explore multiple execution paths based on different possible inputs. After adding these symbolic variables, we compiled each program into LLVM bitcode using Clang. With the bitcode ready, we ran KLEE, which produced several essential output files. The \texttt{.ktest} files contained concrete input values that triggered each path explored. The \texttt{.smt2} files described the symbolic constraints in a format external SMT solvers can understand. KLEE also generated \texttt{.kquery} files, which represent constraints in KLEE's internal format. Finally, the \texttt{.istats} files provided performance metrics like how long they took and how much of the code was covered. This gave us a better sense of how efficiently symbolic execution was performed.


The dataset consists of 100 C programs, each averaging around 15.62 lines of code. Even though these programs were fairly short, running symbolic execution with KLEE on them resulted in surprisingly many outputs. On average, each program generated around 25.48 files for the three key output types: \texttt{.kquery}, \texttt{.smt2}, and \texttt{.ktest}. This shows that even small programs can lead to a high number of symbolic paths, mainly due to branching conditions and logical complexity in the code. It's also important to point out that KLEE always creates the same number of \texttt{.kquery}, \texttt{.smt2}, and \texttt{.ktest} files—one of each per execution path—so the number of these files directly reflects how many paths KLEE explored.


\subsection{Identifying the Most constrained execution path}

We used \texttt{.kquery} files to identify the most constrained execution path because these files directly show how KLEE internally represents constraints, giving us a clear and accurate view of path complexity. Compared to \texttt{.smt} files, \texttt{.kquery} files offer a simpler and more direct way to understand the constraints involved. The path with the most constraints naturally indicates the most complex scenario in the program since it requires checking a more significant number of conditions. To determine the complexity, we counted the total number of conditions or constraints and similar operations in each \texttt{.kquery} file, selecting the file with the highest count as the most complex path for our analysis. The average complexity per \texttt{.kquery file} across all programs in our dataset was 11.39, indicating that most execution paths contained a relatively modest number of constraints. However, our automated script identified one program that stood out significantly—KLEE generated 180 \texttt{.kquery} files for it, and the most constrained file reached a complexity score of 182 despite the program having only 13 lines of source code. This demonstrates how even small programs can produce highly complex execution paths during symbolic execution.


\section{Experiment}
We used the GPT-4o model for our experiments to explore whether it could generate KLEE-like outputs from C source code. We started by collecting KLEE outputs from 100 different C programs. We split the dataset into 80 programs for training and 20 programs for testing. To prepare the data, we structured each example in a conversation-like format from which the model could learn. The input included the original C program, and the expected output consisted of KLEE-generated files—\texttt{.ktest}, \texttt{.smt2}, \texttt{.kquery}, and \texttt{.istats}—along with the most constrained execution path in \texttt{.kquery} format. This format helped the model learn to respond as if it were simulating KLEE. We also wrote a script to automate the formatting process and keep everything consistent, making it easier to fine-tune the model and evaluate its performance on the test set.

\begin{table*}[htbp]
\caption{Comparison Between Actual and Predicted KLEE Output and Critical Path}
\centering
\begin{tabular}{|c|c|c|c|c|c|c|c|c|}
\hline
\textbf{Program} & \textbf{Lines of} & \textbf{Average} 
& \textbf{Actual} & \textbf{Predicted} & \textbf{\#Files}
& \textbf{Most Constraint Path} & \textbf{Correct Path} & \textbf{Most Constraint Path} 
 \\

\textbf{No.}&\textbf{code (LOC)} & \textbf{Complexity} 
& \textbf{\#Files} &\textbf{\#Files} &\textbf{ Match?}
&\textbf{Complexity (Actual)} & \textbf{Predicted?} &\textbf{Complexity (Predicated)}  
\\

\hline
1  & 14 &12.82 
& 11   & 9  & N
& 22  & N & N/A  \\
\hline
2  & 15 & 5.60  & 5    & 8  & N 
& 8   & N & N/A \\
\hline
3  & 23 &12.00  & 22   & 19 & N 
& 15  & N & N/A\\
\hline
4  & 25 &2.67 & 3    & 3  & Y 
& 4   & N & N/A  \\
\hline
5  & 19 &9.00  & 10   & 8  & N 
& 14  & N & N/A \\
\hline
6  & 11 &4.25  & 4    & 5  & N 
& 6   & N & N/A \\
\hline
7  & 22 &5.14  & 7    & 6  & N 
& 8   & N & N/A \\
\hline
8  & 16 &27.94  & 51   & 28 & N 
& 52  & N & N/A \\
\hline
9  & 14 &11.29  & 21   & 6  & N 
& 40  & N & N/A \\
\hline
10  & 14 &6.40  & 10   & 16 & N 
& 18  & N & N/A \\
\hline
11  & 10 &3.75    & 4    & 4  & Y 
& 5   & Y & 6 \\
\hline
12  & 15 &61.66 & 1002 & 12 & N 
& 114 & N & N/A  \\
\hline
13  & 13 &93.47  & 180  & 21 & N 
& 182 & N & N/A \\
\hline
14  & 9  &13.36   & 22   & 23 & N 
& 23  & Y & 23 \\
\hline
15  & 9 &13.36   & 22   & 22 & Y 
& 23  & Y & 22 \\
\hline
16  & 7 &2.33  & 3    & 27 & N 
& 4   & N & N/A \\
\hline
17  & 16 &15.18 & 61   & 21 & N 
& 26  & N & N/A  \\
\hline
18  & 11 &16.58  & 12   & 11 & N 
& 31  & N & N/A \\
\hline
19  & 11 &14.12  & 41   & 27 & N 
& 23  & N & N/A \\
\hline
20 & 11 &4.40    & 5    & 5  & Y 
& 6   & Y & 6 \\
\hline

\multicolumn{3}{|r|}{} &\multicolumn{3}{c|}{\textbf{20\% accuracy in file generation }}
& \multicolumn{3}{|c|}{\textbf{20\% accuracy in most constrained path prediction.}} 
\\

\hline
\end{tabular}
\label{tab:accuracy-summary}
\end{table*}

Table \ref{tab:accuracy-summary} shows a side-by-side comparison of the actual outputs generated by KLEE and the outputs predicted by our fine-tuned GPT-4o model for 20 test programs. Each row represents one program, and the columns highlight different metrics we used to evaluate how well the model performed—such as complexity scores, path prediction accuracy, and the number of generated files.

\noindent \textbf{Program No.} is just the ID we assigned to each program in our test set. We tested 20 different programs in total.\\
    
\noindent \textbf{Lines of Code} show the size of each program. Most of the programs are relatively short, ranging from 7 to 25 lines. This kept the experiments manageable while giving us enough logic and structure variety to evaluate the model correctly. \\
    
\noindent \textbf{Average Complexity} tells us how many constraint operations appeared on average in the .kquery files KLEE produced for each program. This gives a rough estimate of how complex the execution paths were. Some programs had just a few simple conditions, while others had many, making their paths much harder to analyze. \\

\noindent \textbf{Actual \#Files} shows how many .kquery, .ktest, and .smt2 files KLEE generated for each program. These numbers can vary a lot from one program to another because they depend on how many different execution paths KLEE explored. Programs with more conditions and branches usually lead to more paths, which means more output files. \\

\noindent \textbf{Predicted \#Files} shows how many files the fine-tuned model thought should be generated for each program. In most cases, these numbers didn’t match what KLEE produced. The results were inconsistent, and the model often overestimated or underestimated the file counts. This shows that the model struggled to understand how many execution paths a program would create and couldn’t accurately predict the number of .kquery, .ktest, and .smt2 files KLEE would generate.\\

\noindent \textbf{\#Files Match?} Highlights whether the model could precisely match the number of output files generated by KLEE. With only 20\% accuracy, it’s clear the model struggles with producing exact symbolic execution artifacts. This outcome points to a broader challenge—LLMs may recognize common code patterns, but symbolic execution requires exact path enumeration involving logical precision and constraint tracking. Unlike traditional tools like KLEE that operate on formal semantics, the model relies on statistical patterns from training data, which limits its ability to replicate deterministic results. So, while the predictions might look structurally reasonable, they often miss the underlying path logic needed to generate an accurate file count. \\
    
\noindent \textbf{Most Constrained Path Complexity (Actual)} shows how many constraints appeared in the hardest .kquery file for each program, giving us a clear idea of how complex that path was. For example, Program 12, which had only 15 lines of code, had a complexity score of 114, while Program 13, with just 13 lines, reached a score of 182. These numbers show that even short programs can lead to complex symbolic paths when they contain many conditions or branches. \\
    
\noindent \textbf{Correct Path Predicted?} checks whether the model correctly identified the most constrained execution path—the one with the highest number of symbolic conditions. It succeeded in only 4 out of 20 cases. In most instances, the model didn’t generate the .kquery file containing the most constrained path, making accurate prediction impossible. This suggests that the model struggles with identifying complexity and producing the critical paths needed for deeper analysis. \\
    
\noindent \textbf{Most Constrained Path Complexity (Predicted)} reflects how close the model’s predicted constraint score was to the actual one, but only when it correctly identified the path. In those 4 cases, it matched the complexity well 50\% of the time, showing that while the model rarely found the right path, it estimated the complexity fairly accurately when it did.


.

\section{Related Work}
Recent research in symbolic execution has focused on improving scalability, path exploration, and constraint handling. Tools like KLEE have been foundational in automatically generating test cases and identifying bugs through exhaustive symbolic path exploration \cite{cadar2008klee, cadar2013symbolic}. However, one major challenge remains the path explosion problem, where the number of execution paths grows rapidly as the program's complexity increases \cite{bessler2021metrinome}. Researchers have proposed methods to prioritize or prune paths using heuristics and constraint ranking \cite{kapus2020pending}, and others have introduced symbolic execution techniques that work better with under-constrained conditions \cite{engler2007under}. While our work does not propose a new method to reduce path explosion directly, we aim to bypass it by leveraging LLMs to simulate the output of symbolic execution tools like KLEE. Our work shifts the focus from optimizing execution to learning from it, offering a different perspective on managing symbolic analysis complexity.

Large Language Models (LLMs) have recently been explored for symbolic execution and related program analysis tasks. Some researchers have used LLMs to generate high-coverage test inputs \cite{nguyen2024generate, pan2024multi}, detect vulnerabilities in smart contracts by integrating symbolic execution and LLMs \cite{sen2024sellm, xia2025symgpt}, and enhance symbolic execution with learned guidance \cite{zaharudin2024enhancing, chenllm}. Other studies showed LLMs can assist in bug detection and static analysis by leveraging learned code semantics \cite{li2023assisting, ibrahimzada2023automated}. Our work extends these ideas by training GPT-4o to replicate symbolic execution output directly, as well as identifying the most constrained path
\section{Conclusion}
In this work, we explored whether a large language model like GPT-4o could simulate the outputs of KLEE. We trained and tested the model using 100 C programs and their corresponding KLEE outputs to see if GPT-4o could learn to produce similar results without actually running KLEE. While the model only achieved about 20\% accuracy in file generation and identifying the most constrained execution path, this early attempt shows that LLMs may still have potential in this area. Even though the results weren’t strong, this project gave us a clearer view of how far off current models are from handling tasks that require precise symbolic reasoning. It also showed that using an LLM could reduce the time and resources needed for symbolic execution, which could be helpful for scaling analysis across large codebases.

Looking ahead, there are a few ways we plan to improve. One major step is giving the model more context, like ASTs, control flow graphs, and data flow graphs, which could help it better understand how programs are structured and how data moves. We also want to test other LLMs to see if different architectures perform better than GPT-4o. Finally, we’ll explore whether the model-generated outputs are good enough to use for real program analysis tasks like bug finding or performance tuning—comparing them directly with KLEE’s results to see how well they hold up.

\bibliographystyle{IEEEtran}
\bibliography{main}

\end{document}